\documentstyle[12pt,epsf]{article}

\newcommand{\be}{\begin{equation}}
\newcommand{\ee}{\end{equation}}
\newcommand{\bea}{\begin{eqnarray}}
\newcommand{\eea}{\end{eqnarray}}
\newcommand{\ba}{\begin{array}}
\newcommand{\ea}{\end{array}}

\newcommand{\rv}{\vec{r}}
\newcommand{\w}{\vec{\omega}}
\newcommand{\del}{\partial}

\newcommand{\journal}[4]{{\rm #1} {\bf #2} (19#3) #4}
\newcommand{\NP}{\journal{Nucl. Phys.}}
\newcommand{\PL}{\journal{Phys. Lett.}}
\newcommand{\PRL}{\journal{Phys. Rev. Lett.}}
\newcommand{\NPPS}{\journal{Nucl. Phys. Proc. Suppl.}}
\newcommand{\MPCPS}{\journal{Math. Proc. Camb. Phil. Soc.}}

\makeatletter
  
  \@addtoreset{equation}{section}
\makeatother

\begin{document}

%
%
\begin{titlepage}

\begin{flushright}
OU-HET 284 \\
hep-th/9712029 \\
December 1997
\end{flushright}
\bigskip
\bigskip
\bigskip
\bigskip
\bigskip

\begin{center}
{\Large \bf
On the Baryonic Branch Root of $N=2$ MQCD
}

\bigskip
\bigskip
\bigskip
Toshio Nakatsu, Kazutoshi Ohta and Takashi Yokono\\
\bigskip
{\small \it
Department of Physics,\\
Graduate School of Science, Osaka University,\\
Toyonaka, Osaka 560, JAPAN
}
\end{center}
\bigskip
\bigskip
\bigskip
\bigskip
\bigskip

\begin{abstract}
We investigate the brane exchange in the framework of $N=2$ MQCD
by using a specific family of $M$ fivebrane configurations
relevant to describe the baryonic branch root. An exchange of $M$
fivebranes is realized in the Taub-NUT geometry and controlled by
the moduli parameter of the configurations. This family also
provides two different descriptions of the root. These
descriptions are examined carefully using the Taub-NUT geometry.
It is shown that they have the same baryonic branch and are
shifted each other by the brane exchange.   
\end{abstract}

\end{titlepage}

%
%
\section{Introduction}

Recently, many interesting results about supersymmetric gauge
field theories in various dimensions have been obtained by
analyzing the effective theory on the worldvolume of branes in
superstring theory.  These field theories can be realized by
branes mostly in Type IIA or Type IIB superstring theory, but
particularly an interesting configuration, which describes $N=2$
supersymmetric QCD, has been proposed in \cite{Witten1} within
the framework of $M$-theory. In this construction a mysterious
hyper-elliptic curve, which is used for the description of the
exact solution of the Coulomb branch of $N=2$ supersymmetric QCD, 
appears as a part of a $M$-theory fivebrane. This $M$-theory
fivebrane description of QCD is called as MQCD for short.  

It is pointed out \cite{SW1,SW2,Seiberg,IS} that there exist various dualities between
supersymmetric gauge field theories and these dualities play an
important role for our understanding of their non-perturbative
dynamics. Several steps to clarify an origin of these dualities
from the string theory viewpoint have been taken. For example, the
mirror symmetry in three-dimensional gauge theory besides the
non-Abelian duality in $N=1$ four-dimensional gauge theory
possibly reduce to an exchange of branes in Type IIB or Type IIA
theory \cite{HW,EGK}. In this course of explanation, we need a novel concept
that a brane can be created when two different branes cross each
other. However, since this phenomenon of the crossing is owing to
the strong coupling dynamics of Type IIA or Type IIB theory, it
is still difficult to treat it correctly in these theories.    

On the other hand, $M$-theory fortunately includes the strong
coupling dynamics of Type IIA theory in its semi-classical
description \cite{Witten3}. So one can expect that the ``brane creation''
accompanied by exchanging branes can be understood via
semi-classical analysis of $M$-theory. One of the motivations of
this paper is to understand the process of brane exchange in Type
IIA theory from a point of view of $M$-theory brane
configuration. Our approach has some resemblance to  the field
theoretical approaches to prove the non-Abelian duality by
studying a flow from $N=2$ theory to $N=1$ theory
\cite{LS,APS,HMS}. In these approaches the dual theory is
obtained as the flow from an effective theory at the baryonic
branch root. Here the baryonic branch is one of Higgs branches in
$N=2$ supersymmetric QCD, where the gauge symmetry is completely
broken by the Higgs mechanism and ``baryonic'' fields can have
vevs, and ``root'' means vacua where the Higgs branches meet the
Coulomb branch. We study a specific family of $M$-theory
configurations which realize this baryonic branch root by taking
a suitable scaling limit. It consists of curves of $N=2$ MQCD
fivebranes having a discrete ${\bf Z}_{2N_c-N_f}$ symmetry and
being maximally degenerated. 
We
investigate the brane exchange in this $M$-theory
configurations and will give a
detailed interpretation of the brane creation by exchanging
branes. 

In section 2 we study a realization of $N=2$ supersymmetric theory
by $M$-theory fivebrane configuration. In this description the
worldvolume of fivebrane includes a hyper-elliptic curve, the
so-called Seiberg-Witten curve, as a part. When the underlying
field theory does not have a matter hypermultiplet, this curve
is embedded in the flat space, while if including matter, the curve
becomes one embedded into a multi Taub-NUT space. This embedding
of the curve is studied in detail by using a concrete metric of
the multi Taub-NUT space. We also classify the BPS states in
$N=2$ MQCD with respect to topology of $M$-theory membranes
with minimal area.  

In section 3 we consider the aforementioned configurations of $M$
fivebrane. They are constructed by modifying curves of a finite
(scale-invariant) gauge theory so that fivebranes have definite
positions, which becomes important to explain the exchange of
fivebranes in $M$-theory. Besides this, the bare coupling
constant $\tau$ of the finite gauge theory plays the role of a
modulus of these configurations which controls the asymptotic
position of fivebrane. Namely it is a family of $M$
fivebrane configurations parametrised by $\tau$. Changing the
value of the bare coupling constant and considering the
configuration at each value we find that the brane exchange
actually occurs on a semi-circle with radius one in the upper
half $\tau$-plane. Physical significance of this brane exchange
becomes clear when the underlying field theory is investigated.
On this semi-circle the original configuration ($|\tau|>1$) which
describes the baryonic branch root of the ``electric'' theory
shifts to a dual one. This dual theory ($|\tau|<1$) has solitonic
states of the original theory as elementary massless spectrum. The
brane exchange simultaneously exchanges elementary states and
solitonic states. The dual configuration provides another
description of the baryonic branch root of $N=2$ MQCD. Therefore
one can expect that these two configurations give the $N=1$
non-Abelian dual brane configurations \cite{EGK,SS} after rotating a
part of brane \cite{Barbon,HOO,Witten2}.

%
%
\section{$M$-theory fivebranes, membranes and MQCD spectra}

\subsection{Background space-time geometry}

Four-dimensional gauge theories with $N=2$ supersymmetry can be
realized as effective theories on the world-volume of a
$M$-theory fivebrane.  A different type of gauge theories
requires a different topology of $M$-theory fivebrane and a
different eleven-dimensional background where the fivebrane is
embedded. This realization of $N=2$ supersymmetric QCD via the
world-volume effective theory of the $M$-theory fivebrane is
called $N=2$ $M$-theory QCD (MQCD for short.) MQCD does not
exactly coincide with an ordinary supersymmetric QCD in
four-dimension, but is considered to belong to the same
universality class. Moreover many difficulties appearing in the field
theoretical analysis of the supersymmetric QCD vacua, which are
mainly due to their singularities, are resolved within the
framework of $M$-theory. So, MQCD is a very useful
tool for our understanding of the dynamics of supersymmetric QCD.  

Consider an eleven-dimensional manifold $M^{1,10}$ of
$M$-theory. Let us suppose $M^{1,10}$ admits to have the form 
\be
M^{1,10} \simeq {\bf R}^{1,3} \times X^{7}.
\label{decomposition}
\ee
${\bf R}^{1,3}$ is the four-dimensional space-time where $N=2$
supersymmetric theory will exist. $X^7$ is a (non-compact)
seven-dimensional manifold which suffers several constraints due
to the requirement of $N=2$ supersymmetry on the worldvolume. 
The supersymmetry of $M$-theory in the eleven-dimensions is
generally broken by this product space structure of $M^{1,10}$.
However, if the submanifold $X^7$ has a non-trivial holonomy
group, some of supersymmetries are survived on the four-dimensional
space-time ${\bf R}^{1,3}$. Recall that we ultimately realize
$N=2$ supersymmetry on the worldvolume of a $M$-theory fivebrane,
strictly speaking, on its four-dimensional part which is
identified with ${\bf R}^{1,3}$ in (\ref{decomposition}). The
fivebrane itself will be introduced soon later as a BPS
saturated state which breaks the half of the surviving
supersymmetries. So, with this reason, we must take $X^7$ as a
submanifold which keeps $N=4$ supersymmetry on the
four-dimensional space-time ${\bf R}^{1,3}$. Namely, the holonomy
group of $X^7$ is required to be isomorphic to $SU(2)$. It is the
subgroup of a maximal holonomy group $SO(7)$ for a generic
seven-manifold.  

This requirement for the holonomy group reduces the
seven-manifold $X^7$ to be
\be
X^7 \simeq {\bf R}^3 \times Q^4,
\label{X7}
\ee
where ${\bf R}^3$ is a flat three-manifold and $Q^4$ is a
four-manifold with $SU(2)$ holonomy, that is, a hyper-K\"ahler
manifold. This hyper-K\"ahler manifold should be chosen
appropriately according to whether the theory on ${\bf R}^{1,3}$
contains matter hypermultiplets or not.

\subsection{Pure $N=2$ MQCD}

Let us describe a configuration of $M$-theory fivebrane suitable
to our purpose. Since we would like to leave $N=2$ supersymmetry
on ${\bf R}^{1,3}$ as the supersymmetry of the worldvolume
effective theory of fivebrane, the worldvolume must fill all
of ${\bf R}^{1,3}$. The rest of the fivebrane is a
two-dimensional surface $\Sigma$ in $X^7$.   

The Lorentz group $SO(3)\simeq SU(2)/{\bf Z}_2$ of ${\bf R}^3$
in (\ref{X7}) turns out, by considering its action on
covariantly constant Majorana spinors on ${\bf R}^{1,3}$, to be
the $R$-symmetry of $N=2$ supersymmetry algebra. 
In order to preserve this symmetry $\Sigma$ must lie at a single point in ${\bf R}^3$.
It can only spread in $Q^4$ as a
two-dimensional surface. To summarize, the worldvolume of the
fivebrane  must be $R^{1,3}\times\Sigma$ where $R^{1,3}$ is
identified with the four-dimensional space-time and $\Sigma$ is
a two-dimensional surface embedded in $Q^4$.  

Further restriction on the fivebrane world-volume is that
it must be BPS-saturated in order to preserve a half of
supersymmetries. This is achieved by the minimal
embedding of $\Sigma$ into $Q^4$. The area $A_\Sigma$ of the
surface $\Sigma$ is bounded from the below \cite{FS1,HY,Mikhailov,FS2} 
\be
A_\Sigma \geq \frac{1}{2}\int_\Sigma \omega_\Sigma,
\ee
where $\omega_\Sigma$ is the pull-back of a K\"ahler form
$\omega$ of the hyper-K\"ahler manifold $Q^4$. This inequality is
saturated if and only if $\Sigma$ is holomorphically embedded in
$Q^4$ with fixing an appropriate complex structure. Namely, if we
introduce a complex structure of $Q^4$ and define its coordinates
by two complex parameters $y$ and $v$, the surface $\Sigma$ is a
curve in $Q^4$ defined by a holomorphic function    
\be
F(y,v)=0.
\ee

Let $\Sigma$ be a Riemann surface with genus $N_c-1$. Then there
appear $N_c-1$ massless vector multiplets on ${\bf R}^{1,3}$
\cite{Verlinde}. So, the effective theory on ${\bf R}^{1,3}$ is a
supersymmetric $U(1)^{N_c-1}$ gauge theory. In particular let a
Riemann surface be the Seiberg-Witten curve \cite{SW1,SW2} for the pure
$SU(N_c)$ gauge theory \cite{KLY,AF}. 
\be
y^2-2\prod_{a=1}^{N_c}(v-\phi_a)y+\Lambda^{2N_c}=0.
\label{pure}
\ee
If one regards this as a quadratic equation in $y$, two roots
have the forms  
\be
\left\{
\ba{rcl}
y_+&=&\prod_{a=1}^{N_c}(v-\phi_a)
+\sqrt{\prod_{a=1}^{N_c}(v-\phi_a)^2-\Lambda^{2N_c}},\\
y_-&=&\prod_{a=1}^{N_c}(v-\phi_a)
-\sqrt{\prod_{a=1}^{N_c}(v-\phi_a)^2-\Lambda^{2N_c}},
\ea
\right.
\ee
which are considered as a double-sheeted cover of $v$-plane and
have $N_c$ branch cuts on each sheet. 

\begin{figure}[t]
\epsfysize=5cm \centerline{\epsfbox{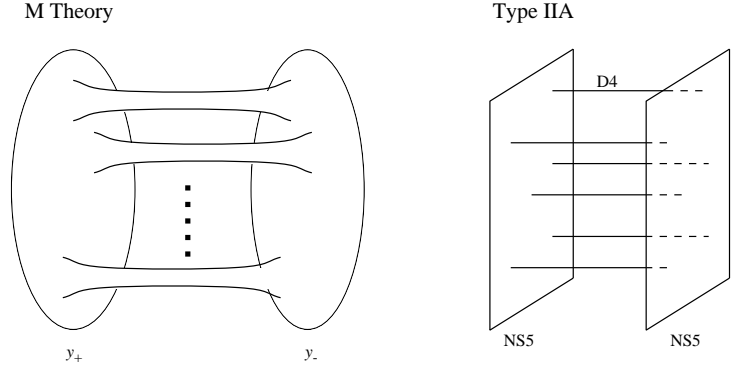}}
\caption{\small
$M$-theory fivebrane and configuration in Type IIA theory.
}
\label{MandIIA}
\end{figure}

We now take, as the four-manifold $Q^4$, a flat space
$Q_0\simeq{\bf R}^3\times S^1$ with coordinates
$(v,b,\sigma)=(x^4+ix^5,x^6,x^{10})/R$, where $R$ is a
compactified radius of $M$-theory. These two sheets are embedded
\cite{Witten1} by the relation $y=\exp(-b-i\sigma)$ into ${\bf
R}^3\times S^1$, connected by the branch cuts. The stretching and
connecting parts of fivebrane wrap once around the circle of the
eleventh-dimension, so these parts become $N_c$ D fourbranes in
Type IIA picture. After all, the single fivebrane in $M$-theory
described by curve (\ref{pure}) becomes two NS fivebranes with
worldvolume $(x^0,x^1,x^2,x^3,x^4,x^5)$ and $N_c$ D fourbranes
with worldvolume $(x^0,x^1,x^2,x^3,[x^6])$\footnote{$[x^6]$
describes a finite interval.} stretching between them in $x^6$
direction. See Fig.\ref{MandIIA}.         

\subsection{$N=2$ MQCD with matter}

In Type IIA picture an inclusion of $N_f$ matter hypermultiplets
into the above pure gauge theory can be achieved by considering
$N_f$ D sixbranes with worldvolume
$(x^0,x^1,x^2,x^3,x^7,x^8,x^9)$ and then putting these sixbranes
between two NS fivebranes. In such a configuration there appears
$N=2$ $SU(N_c)$ supersymmetric QCD with $N_f$ flavors on their
common worldvolume $(x^0,x^1,x^2,x^3)$. This is because the open
string sector between $N_c$ D fourbranes and $N_f$ D sixbranes of
the configuration gives hypermultiplets which belong to the
fundamental representations both of the gauge and flavor groups. 
  
D sixbranes could be regarded as the ``Kaluza-Klein monopoles''
of $M$-theory compactified to Type IIA theory with a circle
$S^1$, since they are magnetically charged with respect to the
$U(1)$ gauge field associated with $S^1$. The $N_f$ sixbranes
transmute \cite{Townsend} the flat space $Q_0={\bf R}^3\times
S^1$ into the multi Taub-NUT space $Q$, which is still
hyper-K\"ahler manifold. Since the sixbranes play the role of
matter hypermultiplets, the Seiberg-Witten curve which is a part
of the fivebrane in $M$-theory changes to the curve of $N=2$
supersymmetric QCD with $N_f$ flavors \cite{HO,APSh}
\be
y^2-2\prod_{a=1}^{N_c}(v-\phi_a)y
+\Lambda^{2N_c-N_f}\prod_{i=1}^{N_f}(v-e_i)=0,
\label{with matter}
\ee
where $e_i$ are the bare masses of the matter hypermultiplets
and identified with the positions of the sixbranes in $v$-plane.
Now the above curve is embedded in the multi Taub-NUT space $Q$. 

To provide a detailed description of the embedding of
curve (\ref{with matter}) into the multi Taub-NUT space $Q$, we
will first deal with the multi Taub-NUT metric. The multi
Taub-NUT metric has the following standard form
\cite{Hawking} 
\be
ds^2=\frac{V}{4}d{\rv}^2+\frac{1}{4V}(d\sigma+\w\cdot d\rv)^2,
\label{mTN}
\ee
where we introduce the coordinates $\rv$ and $\sigma$
of the four-dimensional space $Q$. The potential $V$ is given by  
\be
V=1+\sum_{i=1}^{N_f}\frac{1}{|\rv-\rv_i|},
\ee
where $\rv_i$ represents the position of the $i$-th
sixbrane. The $U(1)$ gauge field $A=\w\cdot d\rv$ is determined
by the relation
\be
\vec{\nabla}\times\w=\vec{\nabla}V.
\ee

Since $Q$ is a hyper-K\"ahler manifold it will have a complex
structure which fits curve (\ref{with matter}). With such a
complex structure one may expect that the multi Taub-NUT metric
becomes K\"ahler. In order to describe it, let us separate $\rv$
into two parts, $b\in{\bf R}$ and $v\in{\bf C}$. Using
these variables, metric 
(\ref{mTN}) acquires the form \cite{NOYY1}
\footnote{
We derive expression (\ref{Kahler metric}) of the multi Taub-NUT
metric by applying the technique investigated by Hitchin
\cite{Hitchin}. In appendix A, we present the derivation to make
this article a self-contained one.} 
\be
ds^2=\frac{V}{4}dvd\bar{v}
+\frac{1}{4V}\left(\frac{2dy}{y}-\delta dv\right)
\overline{\left(\frac{2dy}{y}-\delta dv\right)}
\label{Kahler metric}
\ee
with
\bea
V &=&
1+\sum_{i=1}^{N_f}\frac{1}{\Delta_i},\\
y &=&
C e^{-(b+i\sigma)/2}
\prod_{i=1}^{N_f}(-b+b_i+\Delta_i)^{1/2},
\label{y in mTN}\\
\delta
&=& \sum_{i=1}^{N_f}
\frac{1}{\Delta_i}\frac{b-b_i+\Delta_i}{v-e_i},\\
\Delta_i
&=& \sqrt{(b-b_i)^2+|v-e_i|^2},
\eea
where $C$ is a constant and $(b_i,e_i)$ represents again
the position of the $i$-th sixbrane in this coordinate.
While $y$ in eq.(\ref{y in mTN}) is determined rather explicitly 
by $v,b$ and $\sigma$,
one can also regard that $v$ and $y$ give the holomorphic
coordinates of $Q$.
With this complex structure the multi Taub-NUT metric
becomes K\"ahler as is clear from (\ref{Kahler metric}).
The K\"ahler form $\omega$ is given by
\be
\omega=i\frac{V}{4}dv\wedge d\bar{v}
+\frac{i}{4V}\left(\frac{2dy}{y}-\delta dv\right)\wedge
\overline{\left(\frac{2dy}{y}-\delta dv\right)},
\label{Kahler form}
\ee
and the holomorphic two-form $\Omega$ becomes
\be
\Omega=\frac{i}{2}dv\wedge\frac{dy}{y},
\label{holomorphic two-form}
\ee
which satisfies the relation,
$\frac{1}{2}\omega\wedge\omega=\Omega\wedge\bar{\Omega}$.

Instead of $y$ in (\ref{y in mTN}) one can also take another
choice of the holomorphic coordinates of $Q$ which corresponds to
another branch of curve (\ref{with matter}). It can be given by
the holomorphic coordinates $(v,z)$, where $z$ is introduced by
\be
z =
C' e^{(b+i\sigma)/2}
\prod_{i=1}^{N_f}
\frac{v-e_i}{|v-e_i|}
(b-b_i+\Delta_i)^{1/2}.
\label{z in mTN}
\ee
Notice that $z$ is related to $y$ by the relation
\be
yz=\Lambda^{2N_c-N_f}\prod_{i=1}^{N_f}(v-e_i).
\label{yz=v}
\ee

Eq. (\ref{yz=v}) shows that $Q$ describes a resolution of
$A_{N_f-1}$ simple singularity, $yz=v^{N_f}$. It is resolved by a
chain of $(N_f-1)$ holomorphic two-cycles. Each two-cycle intersects
the next at the position of one of these $N_f$ sixbranes. Let us
denote a two-cycle between two sixbranes $(e_i,b_i)$ and
$(e_{i+1},b_{i+1})$ by $C_i$. Integrals of the two-forms
$\omega$ and $\Omega$ on $C_i$ become \cite{NOYY1}
\bea
\int_{C_i}\Omega &=&  \pi (e_{i+1}-e_{i}),\\
\int_{C_i}\omega &=& 2\pi (b_{i+1}-b_{i}),
\eea
which give the difference between the positions of these two
sixbranes\footnote{These periods are derived by following
\cite{Hitchin}. In appendix B, we summarize them.}.

\begin{figure}[t]
\epsfysize=5cm \centerline{\epsfbox{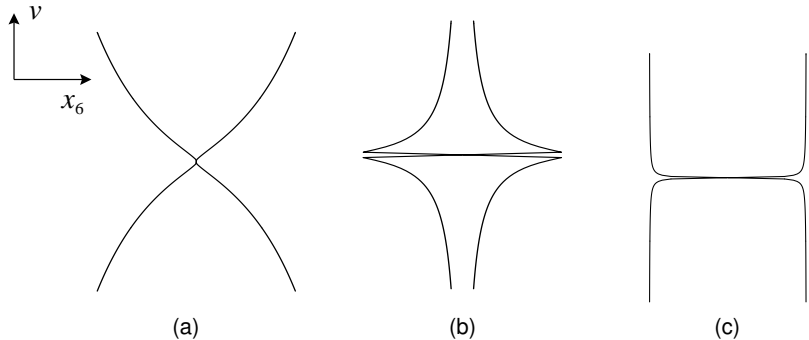}}
\caption{\small
Sections of the Seiberg-Witten curves $\Sigma$ in $M$ theory.
(a) is for asymptotically free theory ($N_f < 2N_c$),
(b) is for IR free theory ($N_f > 2N_c$) and
(c) is for finite theory ($N_f = 2N_c$).
The distance of two NS fivebranes at large $v$ is
infinite, zero and finite, respectively.
}
\label{curveInTN}
\end{figure}

The curve $\Sigma$ described by eq.(\ref{with matter}) is embedded
into $Q$ by using the holomorphic coordinates $(v,y)$ given
above. Since $y$ is related with $b$ and $\sigma$ by eq.(\ref{y
in mTN}) it makes possible to describe the embedding of the curve
$\Sigma$, say, in $(v,b)$-space. Some cases are sketched in
Fig.\ref{curveInTN}. 

\subsection{Membranes and BPS states}

In this subsection we classify the BPS states of $N=2$ MQCD
according to the topology of membrane.    

Let us consider a membrane which worldvolume is ${\bf R}\times
D$, where a two-dimensional surface $D$ is embedded into $Q$ with
its boundary $C=\del D$ lying on the curve $\Sigma$. Area of the
membrane, which is proportional to the membrane mass, satisfies
the inequality \cite{FS1,HY,Mikhailov,FS2} 
\be
A_D\geq\frac{1}{2}\left|\int_D\Omega_D\right|,
\label{membrane area}
\ee
where $\Omega_D$ is the pull-back to $D$ of the holomorphic
two-form (\ref{holomorphic two-form}).  $\Omega_D$ can be written
as an exact form on $D$, that is, $\Omega_D=d\lambda$ where
$\lambda=\frac{i}{2}v\frac{dy}{y}$. Then the integral in the
r.h.s. of inequality (\ref{membrane area}) can be evaluated as a
boundary integral  
\be
\int_D\Omega_D=\oint_C\lambda.
\label{mass formula}
\ee
This boundary integral is nothing but the Seiberg-Witten mass
formula for four-dimensional $N=2$ theories. In particular
$\lambda$ can be regarded as the Seiberg-Witten differential \cite{SW1,SW2}.
Thus the membrane of minimal area gives the BPS saturated state
of $N=2$ MQCD with its mass being its area up to the membrane
tension.  

Quantum numbers of the BPS saturated state or the membrane of
minimal area can be read from homology class $[C]$ of its
boundary in $H_1(\Sigma,{\bf Z})$. We will examine some simple
examples of these BPS saturated membranes. 

Gauge fields and W bosons appear as Type IIA strings stretching
between same or different D fourbranes. In $M$-theory, since IIA
string corresponds to a membrane wrapping around a circle of the
compactified eleven-dimension, gauge fields and W bosons can be
identified cylindrical membranes connecting between cycles
$\alpha_i$ on the curve $\Sigma$. $\alpha_i$ is a closed path
surrounding the $i$-th branch cut on the Riemann sheet of $\Sigma$.
If two cycles are coincident the corresponding membrane of
minimal area represents gauge boson, otherwise W boson. Both
states are electrically charged and integrals (\ref{mass formula})
on the cycles $\alpha_i$ give their correct BPS masses.  

\begin{figure}[t]
\epsfysize=5cm \centerline{\epsfbox{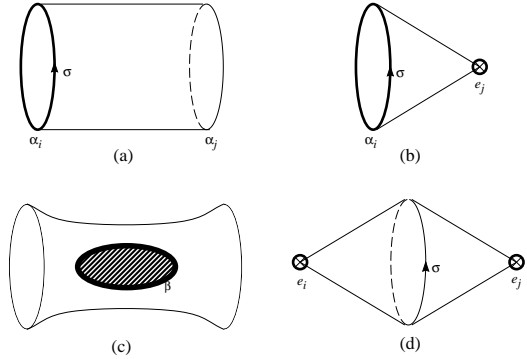}}
\caption{\small
Topology of the minimal surfaces in $Q$ corresponding to the
BPS saturated states in $N=2$ MQCD. (a) electrically charged
gauge boson or W boson in vector multiplet. (b) electrically
charged quark with bare mass $e_j$ in hypermultiplet. (c)
magnetically charged monopole in hypermultiplet. (d) gauge
singlet quark - anti-quark bound state. (``meson''.) 
}
\end{figure}

Electrically charged matter (quark) hypermultiplets are Type IIA
strings between D fourbranes and D sixbranes. In $M$-theory they
are the membranes connecting the cycles $\alpha_i$ and the
sixbranes. Topology of these membranes is a disk with a puncture
at the position of the sixbrane. It is like a cone. Since the
Seiberg-Witten differential $\lambda$ has a pole at $v=e_j$ with
its residue equal to the bare mass $e_j$, the BPS mass of the
membrane is sum of the period along the cycle $\alpha_i$ and
the bare mass $e_j$. This also agree with the result of
\cite{SW2}.   

Monopole is a magnetically charged hypermultiplet. Therefore the
boundary of the corresponding membrane must be on a cycle
$\beta_i$ dual to $\alpha_i$. Topology of this membrane is a
disk. Therefore integral (\ref{mass formula}) gives the correct
monopole mass.   

Finally, let us consider a slight curious state in $N=2$ MQCD.
Suppose that two membranes representing quark hypermultiplets
have their common boundary on the $\alpha$-cycles of $\Sigma$. If
one paste these two membranes along their common boundary, one
can obtain another membrane which have a sphere topology with two
punctures at the positions of the sixbranes. This state is not
charged under the gauge group. This gauge singlet
state will represent the quark and anti-quark bound state, that is,
``meson'' hypermultiplet $M^i_j=Q^i_a\tilde{Q}^a_j$. However,
since this membrane does not end on the fivebrane, the state does
not appear in the worldvolume effective theory on ${\bf
R}^{1,3}$. So we must add the worldvolume ${\bf R}^{1,3}$ to the
``meson'' and regard it as a fivebrane with worldvolume ${\bf
R}^{1,3}\times S^2$, where $S^2$ is a two-sphere in the multi
Taub-NUT space $Q$.  

This two-sphere in $Q$ can be identified with the two-cycle which
resolve the simple singularity. When the theory enter the Higgs
branch, a part of the fivebrane described by the curve begins to
wrap the two-cycle and divides into ``meson'' parts
\cite{Witten1}. According to the analysis of the embedding of the
curve into $Q$, the number of these fivebranes wrapping the
two-cycles is exactly equal to the dimensions of the Higgs branch
\cite{HOO,NOYY1}. Therefore the above ``meson'' variables
parameterize the moduli space of the Higgs branch.

%
%
\section{Baryonic branch root of MQCD}

In this section we will examine the root of  baryonic branch from
the MQCD view-point. The baryonic branch and the Coulomb branch
encounter with each other at this root. Field theoretical
analysis \cite{APS} shows that the baryonic branch root is a
single point where the underlying theory is invariant under the
${\bf Z}_{2N_c-N_f}$ discrete symmetry (anomaly-free subgroup of
the classical $U(1)_R$ symmetry). Though the $SU(N_c)$ gauge
symmetry is broken to $U(1)^{N_c-1}$ at a generic point of the
Coulomb branch, $SU(N_f-N_c)\times U(1)^{2N_c-N_f}$ non-Abelian
gauge symmetry is allowed at this root and the corresponding
gauge theory can be regarded as a IR-effective theory of the
root. Moreover, this IR-free $SU(N_f-N_c)\times U(1)^{2N_c-N_f}$
gauge theory has $2N_c-N_f$ massless singlet hypermultiplets
charged only by the $U(1)$ factors in addition to $N_f$ quark
hypermultiplets which belong to the fundamental representations
of $SU(N_f-N_c)$. A baryonic branch of this IR-effective theory
exactly coincides with the baryonic branch of the original
microscopic $SU(N_c)$ theory. Original microscopic theory and
IR-effective theory will be called respectively as ``electric''
theory and ``magnetic'' theory.   

If one flows the IR-effective theory of the baryonic branch root
from $N=2$ to $N=1$ by giving a mass to the adjoint scalar field,
we naively expect to obtain $N=1$ $SU(N_f-N_c)$ gauge theory,
which is a non-Abelian dual to the original gauge theory. In
terms of brane configuration this flow can be interpreted as a
rotation of a part of fivebrane \cite{Barbon,HOO,Witten2}. While
this observation, the non-Abelian duality of $N=1$ gauge
theories can be explained as the exchange of a NS fivebrane and
rotated one in the $N=1$ brane configuration of Type IIA theory
\cite{EGK} and $M$-theory\cite{SS}. 

If these two operations are reversible, we may first exchange
fivebranes in $M$-theory configuration and expect the same result
after the rotation. Since the brane exchange could be related
with the strong coupling dynamics of string theory, to consider
it first in the Taub-NUT geometry, which is rather tractable
than the Calabi-Yau threefold, will provide new insight on the
duality. In order to exchange fivebranes in the $M$-theory
configuration, we need their exact positions in the multi
Taub-NUT space. As we can see in Fig.\ref{curveInTN}, for the AF
theory and IR-free theory the corresponding fivebrane has no
definite position, but for the finite theory the position of
fivebrane asymptotically approaches to a definite value since the
coupling constant of the theory does not run. So we may utilize a
curve of the finite theory in our description of the brane
exchange in $M$-theory.

\subsection{Configuration of fivebrane inspired by finite
theory curve} 

Let us first recall that the Seiberg-Witten curve of $N_f=2N_c$
finite theory has the form \cite{APSh,APS} 
\be
y^2-2\prod_{a=1}^{N_c}(v-\phi_a)y
-h(h+2)
\prod_{j=1}^{2N_c}(v-he_S-e_j)=0,
\label{finite curve}
\ee
where $e_S=\frac{1}{2N_c}\sum e_j$
is the center of the bare masses $e_j$.
$h$ is a specific modular function of the bare coupling
constant $\tau=\frac{\theta}{\pi}+\frac{8\pi i}{g^2}$ and is given
by 
\be
h(\tau)
= \frac{2 \lambda(\tau)}{1-2 \lambda(\tau)}.
\ee
Notice that $\lambda(\tau)$ is the following automorphic function
\be
\lambda(\tau)=
16 q \prod_{n=1}^{\infty}
\left(
\frac{1+q^{2n}}{1+q^{2n-1}}
\right)^8
\ee
with $q=e^{\pi i \tau}$.
The modular transformation of $\lambda(\tau)$ are \cite{WW}
\bea
T &:&  
\lambda(\tau+1)=\frac{\lambda(\tau)}{\lambda(\tau)-1}, \\
S &:&
\lambda(-1/\tau)
=
1- \lambda(\tau),
\eea
which implies that $\lambda(\tau)$ is invariant under the action
of the congruence subgroup of level 2 
\be
\Gamma(2)=
\left\{
\left(
\left.
\begin{array}{cc}
a & b \\
c & d
\end{array}
\right) \in SL(2, {\bf Z})~
\right|~
b,c ~~\mbox{even}
\right\}.
\ee

\begin{figure}[t]
\epsfysize=5cm \centerline{\epsfbox{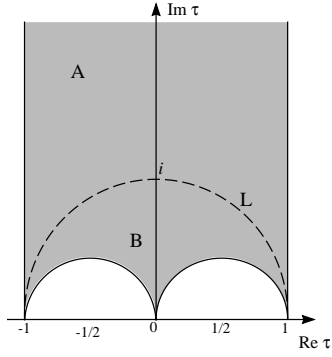}}
\caption{\small
Fundamental domain ${\cal F}$ for $\Gamma(2)$.
}
\label{fundamental domain}
\end{figure}

Let ${\cal F}$ be a fundamental domain for $\Gamma(2)$ depicted
in Fig.\ref{fundamental domain}. The function $\lambda(\tau)$
maps the set ${\cal F}$, one-to-one, onto ${\bf
C}\cup\{\infty\}$. Since $\lambda(\tau)$ satisfies the relation
$\overline{\lambda(\tau)}=\lambda(-\bar{\tau})$, it follows that 
${\rm Re}~\lambda(\tau)=1/2$ for ${}^\forall\tau\in L\equiv
{\cal F}\cap\{~\tau~|~|\tau|=1\}$. Moreover, one can find that 
\be
\left\{
\ba{ccc}
{\rm Re}~\lambda(\tau)<1/2 & {\rm in} & A\equiv {\cal F}\cap\{~\tau~|~|\tau|>1\},\\
{\rm Re}~\lambda(\tau)>1/2 & {\rm in} & B\equiv {\cal F}\cap\{~\tau~|~|\tau|<1\}.
\ea
\right.
\ee
These properties of $\lambda(\tau)$ will be important for
later discussions.

As a result, $h(\tau)$ satisfies the modular property
$h\rightarrow -(h+2)$ under $\tau\rightarrow -1/\tau$. Therefore,
the Seiberg-Witten curve (\ref{finite curve}), which describes
the Coulomb branch of the theory, is invariant under
$S$-transformation if the bare masses admit to be transformed as  
\be
\left\{
\ba{lcl}
e_j &\rightarrow& e_j-2e_S,\\
e_S &\rightarrow& -e_S.
\ea
\right.
\label{D6v}
\ee
With these transformations the Coulomb branch acquires a $SL(2,{\bf Z})$ symmetry. We may call it the $SL(2,{\bf Z})_V$ symmetry since the Coulomb branch is the moduli space of vector multiplets.


Now let us examine a family of the baryonic branch root
embedded in the finite theory. Since the baryonic branch has the
${\bf Z}_{2N_c-N_f}$ symmetry, we require first the finite
theory curve to have this symmetry. By this requirement the vevs
of the adjoint scalar field and the bare masses must be of the
form   
\bea
\phi_a &=&
(0,\ldots,0,\varphi\omega,\varphi\omega^2,\ldots,\varphi\omega^{2N_c-N_f}),\\
e_j &=&
(\underbrace{0,\ldots,0}_{N_f-N_c},m\omega,m\omega^2,\ldots,m\omega^{2N_c-N_f}),
\eea
where $\omega^{2N_c-N_f}=1$. Then, curve (\ref{finite curve}) becomes
\be
y^2-2v^{N_f-N_c}(v^{2N_c-N_f}-\varphi^{2N_c-N_f})y
-h(h+2)
v^{N_f}(v^{2N_c-N_f}-m^{2N_c-N_f})=0.
\label{sym curve}
\ee

Further requirement on the curve is that it should be maximally
degenerated at the baryonic branch root. Namely, all cycles on
the curve must vanish. To establish this, we first rewrite curve
(\ref{sym curve}) in a quadratic form of
$Y=y-\prod_{a=1}^{N_c}(v-\phi_a)$  
\be
Y^2=v^{2(N_f-N_c)}
\left\{(v^{2N_c-N_f}-\varphi^{2N_c-N_f})^2
+h(h+2)v^{2N_c-N_f}(v^{2N_c-N_f}-m^{2N_c-N_f})
\right\}.
\label{YY}
\ee
When the r.h.s. of (\ref{YY}) becomes a perfect square,
all the branches disappear and therefore all the cycles vanish.
The Riemann sheets of the hyperelliptic curve are decoupled to two
complex planes. This situation can occur if and only if the
relation  
\be
m^{2N_c-N_f}=\frac{2}{-h(\tau)}\varphi^{2N_c-N_f},
\label{m ele}
\ee
or
\be
m^{2N_c-N_f}=\frac{2}{h(\tau)+2}\varphi^{2N_c-N_f}
\label{m mag}
\ee
is satisfied. Note that these two relations are equivalent in the
sense that modular transformation $\tau\rightarrow -1/\tau$
exchanges (\ref{m ele}) and (\ref{m mag}). So it is enough to
study only the former without loss of generality. However, this
$S$-transformation of the position of sixbranes does not satisfy
(\ref{D6v}). So it breaks the $SL(2,{\bf Z})_V$ symmetry, which
means that configuration (\ref{m ele}) does not belong to the
category of finite theory curve, but it is still meaningful as a
$M$-theory brane configuration and, after taking a suitable
scaling limit, it turns out to describe the baryonic branch root.  

When relation (\ref{m ele}) is satisfied the curve becomes the
perfect square  
\be
Y^2=v^{2(N_f-N_c)}
\left\{(h+1)v^{2N_c-N_f}+\varphi^{2N_c-N_f}\right\}^2.
\ee
Or equivalently two roots of (\ref{sym curve}) with respect to
$y$ describe two independent complex planes without any branch  
\be
\left\{
\ba{rcl}
y_+ &=& (h+2)v^{N_c},\\
y_- &=& -v^{N_f-N_c}
\left\{hv^{2N_c-N_f}+2\varphi^{2N_c-N_f}\right\}.
\ea
\right.
\label{two planes}
\ee
It means that the single fivebrane embedded in the multi Taub-NUT
space now decouples to two parts. As we discuss subsequently
these two fivebranes admit to have an intersection in the Taub-NUT
space. Though the intersection itself has its origin in the
branch cuts of the Riemann sheets, its behavior for a given
$\tau$, if one consider it in the Taub-NUT space, is quite
different depending on the value of $\tau$.  

\subsection{Intersection of fivebranes in the Taub-NUT space}

We examine their intersection in the Taub-NUT space in detail.
Recall the embeddings of two fivebranes (\ref{two planes}) are
given by the equations 
\be
\left\{
\ba{rcl}
y_+&=&C e^{-(b_++i\sigma_+)/2}\prod^{2N_c}_{j=1}(-b_++b_i+\Delta_i),\\
y_-&=&C e^{-(b_-+i\sigma_-)/2}\prod^{2N_c}_{j=1}(-b_-+b_i+\Delta_i).
\ea
\right.
\ee
Their intersection in the Taub-NUT space can be handled by the
equation $b_+(v)=b_-(v)$, which implies 
\be
\left|\frac{y_+}{y_-}\right|=1.
\label{intersection1}
\ee
This can be regarded as an equation for $v$. The allowed values
of $v$, that is, solutions of the equation, describe the
intersection projected to $v$-plane. Inserting explicit
forms (\ref{two planes}) to (\ref{intersection1}), after a little
calculation, we obtain    
\be
\left|1-\frac{M}{V}\right|=\left|1-\frac{1}{\lambda(\tau)}\right|,
\label{intersection2}
\ee
where $V=v^{2N_c-N_f}$ and $M=m^{2N_c-N_f}$. An implication of
eq.(\ref{intersection2}) may be tractable if one considers it in
$V$-plane rather than $v$-plane. Now $M$ gives the sixbrane
position on $V$-plane. Let us take $M$ as a positive real
quantity and set $V=X+iY$. Then eq.(\ref{intersection2}) becomes   
\be
\left(X-\frac{|\lambda|^2M}{2\gamma-1}\right)^2
+Y^2
=\left|\frac{\lambda(\lambda-1)M}{2\gamma-1}\right|^2,
\label{circle}
\ee
where $\gamma={\rm Re}~\lambda$. If $\gamma\neq 1/2$ it
describes a circle centered at
$X_0\equiv\frac{|\lambda|^2}{2\gamma-1}M$ on $X$-axis with radius
$R\equiv\left|\frac{\lambda(\lambda-1)}{2\gamma-1}\right|M$. Let
us comment some details on this intersection (\ref{circle}). It can be
classified into the following three cases.  

\begin{figure}[t]
\epsfysize=5cm \centerline{\epsfbox{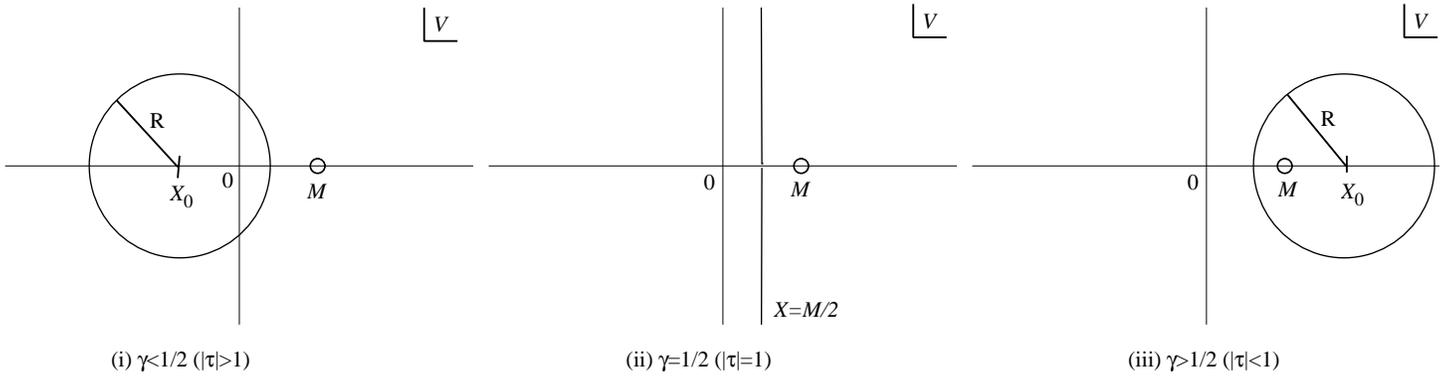}}
\caption{\small
Intersections of two fivebrane on $V$-plane.
}
\label{three cases}
\end{figure}

When $\gamma$ is less than $1/2$ ({\it i.e.} $|\tau|>1$), it
holds that $X_0<0$ and $0<X_0+R<M$. So the origin is inside of
the circle and the sixbrane is located outside of the circle.
When $\gamma$ is equal to $1/2$ ({\it i.e.} $|\tau|=1$),
(\ref{intersection2}) describes a straight line between the
origin and the sixbrane; $X=\frac{1}{2}M$. Finally, when $\gamma$
is greater than $1/2$ ({\it i.e.} $|\tau|<1$), it holds that
$X_0>0$ and $0<X_0-R<M<X_0+R$. So the origin is outside of the
circle while the sixbrane is inside of the circle. They
are depicted in Fig.\ref{three cases}.  

\begin{figure}[t]
\epsfysize=5cm \centerline{\epsfbox{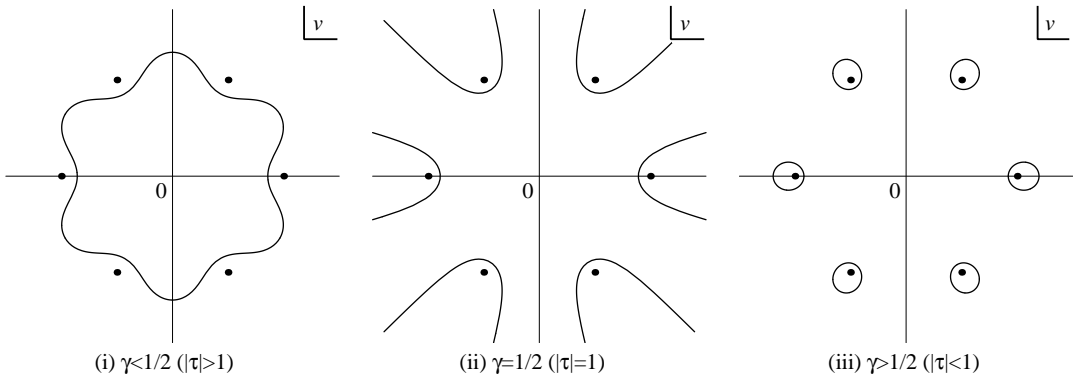}}
\caption{\small
Intersections of two fivebrane on $v$-plane. Dots represent
the positions of the sixbranes. The case of $2N_c-N_f=6$ is
sketched as an example. 
}
\label{v-plane}
\end{figure}

Since $v$ is a ($2N_c-N_f$)-th root of $V$, these circles and
line are copied $2N_c-N_f$ times on the original $v$-plane.
Namely, the circle or line in each case becomes a single wavy
circle which surrounds the origin, $2N_c-N_f$ parabolic lines or
$2N_c-N_f$ circles surrounding each sixbrane, respectively. These
intersections of two fivebranes in the Taub-NUT space are
depicted in Fig.\ref{v-plane}. 

\subsection{Weak coupling limits}

Since the intersection of fivebranes in the Taub-NUT space is
realized in very different fashions depending on the value of
$\tau$, it will bring us different descriptions of the ${\bf
Z}_{2N_c-N_f}$ symmetric and maximally degenerate theory. There
appear two regions of the fundamental domain ${\cal F}$ where
the intersection is qualitatively different from each other.
These two regions are separated by the semi-circle $L$. So, we
can expect, at least, two completely different descriptions of
the theory. 

To determine the massless spectrum of each description, we will
take the weak coupling limit in terms of the bare coupling
constant $\tau$ or its dual $\tilde{\tau}=-1/\tau$. We first
consider the case of the original ``electric'' theory. Let
$m$ in (\ref{m ele}) be a function of $\tau$ with a
fixed constant $\varphi=\Lambda$. In the weak coupling limit
($\tau\rightarrow i\infty$) of the configurations, $2N_c-N_f$
sixbranes are naturally decoupled and curve (\ref{sym curve})
becomes
\be
y^2-2v^{N_f-N_c}(v^{2N_c-N_f}-\Lambda^{2N_c-N_f})y
-4\Lambda^{2N_c-N_f}v^{N_f}=0.
\label{bbroot}
\ee
Note that $\lim_{\tau\rightarrow i\infty} h(\tau)=0$. This
degenerated curve describes the baryonic branch root of the AF
theory with color $SU(N_c)$ and $N_f$ flavor. It exactly agrees
with \cite{APS}. This is nothing but the desired original
``electric'' theory! At this baryonic branch root of the
``electric'' theory, the gauge symmetry is broken to
$SU(N_f-N_c)\times U(1)^{2N_c-N_f}$. Due to the degeneration of
the curve there appear $2N_c-N_f$ mutually local massless
hypermultiplets. Each of them is charged under only one $U(1)$
factor. Thus all massless fields appear as the magnetically
charged solitonic states.    

Next we examine the weak coupling limit of the dual bare coupling
constant $\tilde{\tau}=-1/\tau$. This limit
$\tilde{\tau}\rightarrow i\infty$ corresponds to the limit
$\tau\rightarrow 0$ of the original bare coupling constant. So we
must examine the region $|\tau|<1$. The intersection of two
fivebranes is circles which surround the extra $2N_c-N_f$
sixbranes. The radius of these circles vanish in this limit. Let us
pay attention to these small circles. In the Taub-NUT space they
give rise to two-spheres surrounding the sixbranes and confined
by the two fivebranes. These two-spheres consist of the circles
and the disks in the fivebranes inside their intersection.
Mathematically speaking, what we obtain here is not a two-sphere
but a punctured two-sphere. This is because the sixbranes are
the NUT singularities in the Taub-NUT space and there exist Dirac
strings running from the sixbranes. Two-spheres obtained above
must have intersections with these Dirac strings since they are
surrounding the sixbranes.

\begin{figure}[t]
\epsfysize=5cm \centerline{\epsfbox{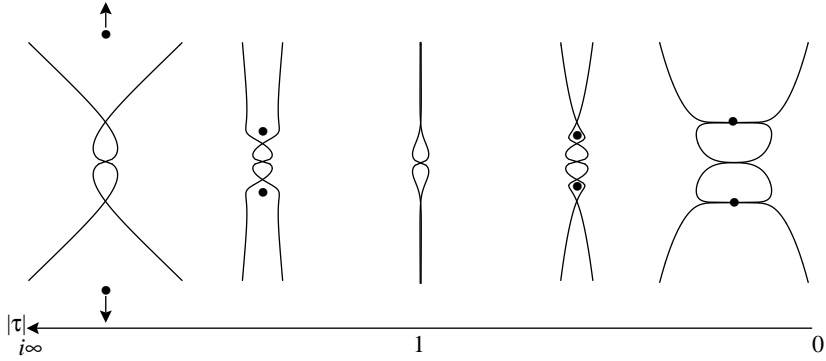}}
\caption{\small
Sections of the intersecting two fivebranes. Dots represent the
sixbranes. A limit to the left ($\tau\rightarrow i\infty$)
describes the baryonic branch root of $SU(N_c)$ theory and a
limit to the right ($\tau\rightarrow 0$) describes the baryonic
branch root of $SU(N_f-N_c)$ theory.   
}
\label{duality}
\end{figure}

On the other hand, if we look at only one side of the
fivebranes, the radius of circles is also considered roughly as
thickness of the fivebrane which stretches from major part lying
at asymptotic position toward the sixbrane. So, in the limit
$\tau\rightarrow 0$ the stretching part of the fivebrane becomes
very fine. Moreover, due to aforementioned puncture on the
fivebrane, this stretching part can be thought to wrap the
eleventh-dimension. Then this part becomes D fourbrane stretching
from NS fivebrane and  touching the D sixbrane in Type IIA
theory\footnote{More detailed analyses are presented in
\cite{NOYY2,Yoshida}.}. (See Fig.\ref{duality}.) 



This observation can be also confirmed from the curve. In the
limit $\tau\rightarrow 0$, it holds $\lim_{\tau\rightarrow
0}m^{2N_c-N_f}(\tau) = \Lambda^{2N_c-N_f}$. So curve
(\ref{sym curve}) becomes 
\be
y^2-2v^{N_f-N_c}(v^{2N_c-N_f}-\Lambda^{2N_c-N_f})y
-h(h+2)
v^{N_f}(v^{2N_c-N_f}-\Lambda^{2N_c-N_f})=0.
\ee
This describes the finite theory whose $2N_c-N_f$ D sixbranes and
D fourbranes are at the same position on $v$-plane and touching
each other. 

The massless spectrum which we can read from the above
limit of the configurations are as follows: $2N_c-N_f$ massless singlet
hypermultiplets obtained from the open string connecting the extra
$2N_c-N_f$ D sixbranes and the D fourbranes touching them.
These hypermultiplets appear now as the elementary states. So
the configuration itself describes the dual ``magnetic'' theory.

In both limits, $\tau\rightarrow i\infty$ and $\tau\rightarrow
0$, same massless spectrum appears, but in different fashions.
Namely they appear as massless solitonic states due to the
monopole singularity or massless elementary states due to the
quark singularity.

\subsection{Brane exchange}

We have seen the massless spectrum of the configuration at the
boundaries of their moduli space, that is, $\tau=i \infty$ and
$\tau=0$. Now consider a path in the moduli space which connects
these two boundaries. In the region $|\tau|>1$ the asymptotic
positions of two fivebranes for large $|v|$ can be read from
(\ref{two planes})   
\be
\left\{
\ba{rcl}
b_+^{as}&=&\ln|h(\tau)+2|,\\
b_-^{as}&=&\ln|-h(\tau)|.
\ea
\right.
\ee
These asymptotic positions coincide with
each other on the semi-circle $L$, since the relative distance
of the asymptotic positions satisfies 
\be
\Delta b^{as }
=\ln\left|\frac{h(\tau)+2}{-h(\tau)}\right|
=\ln\left|\frac{\lambda(\tau)-1}{\lambda(\tau)}\right|=0,
\ee
if $|\tau|=1$. And obviously two positions are exchanged under
the $S$-dual transformation $\tau\rightarrow -1/\tau$. Therefore
one can say that, if $\tau$ moves continuously from one region to
another across the semi-circle $L$, the asymptotic positions of
two fivebranes are passed each other on $L$ and exchanged. 

As we have seen, this brane exchange also exchanges the
solitonic states with the elementary states. Moreover, the
degeneracy of the curve at the origin of $v$-plane, which only
relates to the moduli space of the baryonic branch, is not
affected by this exchange of branes. Therefore the baryonic branches
realized in each region are exactly the same, that is, the
baryonic branch of $SU(N_c)$ MQCD with $N_f$ flavors is
isomorphic to that of $SU(N_f-N_c)$ MQCD with $N_f$ flavors.
Thus, instead of broken $SL(2,{\bf Z})_V$ symmetry, there exists
another $SL(2,{\bf Z})$ symmetry for the baryonic branch, that
is, the moduli space of vacua for the hypermultiplets.  

This situation is very similar to the explanation of Seiberg's
non-Abelian duality in $N=1$ SQCD by the exchange of Type IIA
brane configurations. So if we rotate the above $N=2$ MQCD
configuration and break the supersymmetry to $N=1$, it will give
a proof of the $N=1$ non-Abelian duality via $M$-theory.  

%
%

\bigskip
\bigskip

%
%

\section*{Acknowledgments}

We would like to thank Y. Yoshida 
for useful discussions and and comments. 
T.N. is supported in part by
Grant-in-Aid for Scientific Research 08304001.
K.O. is supported in part by the JSPS
Research Fellowships.

\newpage

%
%
\section*{Appendix}
\appendix
\section{The multi Taub-NUT metric as a K\"ahler metric}

The multi Taub-NUT space is asymptotically flat and looks near infinity
like ${\bf R}^3\times S^1$. If we set the
coordinates $(\rv,\sigma)\equiv(x^1,x^2,x^3,\sigma)\in{\bf
R}^3\times S^1$, where $\sigma$ is periodic
$(0\leq\sigma\leq4\pi)$, the metric is given by 
\cite{Hawking} 
\be
ds^2=\frac{V}{4}d{\rv}^2+\frac{1}{4V}(d\sigma+\w\cdot d\rv)^2,
\label{A:mTN metric}
\ee
where
\be
V=1+\sum_{i=1}^{d}\frac{1}{|\rv-\rv_i|},
\ee
and $\rv_j$ is a position of the $j$-th monopole in ${\bf R}^3$.
$\w$ is the Dirac monopole potential, which satisfies 
\be
\vec{\nabla}\times\w=\vec{\nabla}V.
\label{A:Dirac potential}
\ee
A particular solution of eq.(\ref{A:Dirac potential}) can be chosen as
\be
\left\{
\ba{rcl}
\omega_1&=&0,\\
\omega_2
&=&\sum_{j=1}^{d}
\frac{x^3-x^3_j}{|\rv-\rv_j|(x^1-x^1_j-|\rv-\rv_j|)},\\
\omega_3
&=&\sum_{j=1}^{d}
\frac{-x^2+x^2_j}{|\rv-\rv_j|(x^1-x^1_j-|\rv-\rv_j|)}.
\ea
\right.
\ee
Let us first substitute this solution for $\w\cdot d\rv$ in
metric (\ref{A:mTN metric}), 
\be
\w\cdot d\rv=
{\rm Im}\left(
\sum_{j=1}^{d}\frac{1}{|\rv-\rv_j|}\frac{b-b_j+|\rv-\rv_j|}{v-e_j}dv
\right),
\label{A:wdr}
\ee
where $b\equiv x^1\in{\bf R}$ and $v\equiv x^2+ix^3\in{\bf C}$.
The position of the $j$-th monopole is rewritten in this
coordinates as $\rv_j\mapsto(b_j,e_j)$. We also define the
following quantities for later convenience:
\bea
\Delta_j &\equiv& |\rv-\rv_j|\\
         &=& \sqrt{(b-b_j)^2+|v-e_j|^2},\nonumber\\
\delta &\equiv&
\sum_{j=1}^{d}\frac{1}{\Delta_j}\frac{(b-b_j)+\Delta_j}{v-e_j}.
\eea
Then (\ref{A:wdr}) simply becomes
\be
\w\cdot d\rv={\rm Im}(\delta dv).
\label{A:wdr2}
\ee

Next we introduce the quantity
\be
y\equiv C e^{-(b+i\sigma)/2}\prod_{j=1}^{d}(-b+b_j+\Delta_j)^{1/2},
\label{A:y}
\ee
where $C$ is some constant. After some straightforward
calculation we can find 
\be
\frac{2dy}{y}=-Vdb-id\sigma+{\rm Re}(\delta dv).
\label{A:dlogy}
\ee
Using eq.(\ref{A:dlogy}) the following equalities can be shown:
\bea
\left(\frac{2dy}{y}-\delta dv\right)
\overline{\left(\frac{2dy}{y}-\delta dv\right)}
&=&
\left\{
{\rm Re}\left(
\frac{2dy}{y}-\delta dv
\right)
\right\}^2 
+
\left\{
{\rm Im}\left(
\frac{2dy}{y}-\delta dv
\right)
\right\}^2 \nonumber\\
&=&
V^2db^2+\{d\sigma+{\rm Im}(\delta dv)\}^2.
\label{A:dlny^2}
\eea

Finally, inserting eq.(\ref{A:wdr2}) into metric (\ref{A:mTN
metric}) and rewriting it using eq.(\ref{A:dlny^2}), the multi
Taub-NUT metric acquires the form of a K\"ahler metric: 
\bea
ds^2&=&
\frac{V}{4}(db^2+dvd\bar{v})
+\frac{1}{4V}\{d\sigma+{\rm Im}(\delta dv)\}^2 \\
&=&\frac{V}{4}dvd\bar{v}
+\frac{1}{4V}\left(\frac{2dy}{y}-\delta dv\right)
\overline{\left(\frac{2dy}{y}-\delta dv\right)}.
\eea

\section{Integrals on vanishing two-cycles in the multi Taub-NUT
space} 

A vanishing two cycle $C_j$ between two monopoles at $(b_j,e_j)$
and $(b_{j+1},e_{j+1})$ can be represented in $(b,v)$-space as a
line 
\be
\left\{
\ba{rcl}
b&=&\lambda(b_{j+1}-b_j)+b_j,\\
v&=&\lambda(e_{j+1}-e_j)+e_j,
\ea
\right.
\ee
where $\lambda$ is a real parameter, $0\leq\lambda\leq 1$. Let
$y(\lambda,\sigma)$ be a function defined by restricting $y$ on
$C_j$, that is, $y(\lambda,\sigma)\equiv y|_{C_j}$. Using the
explicit form (\ref{A:y}) of $y$, $y(\lambda,\sigma)$ turns out
to have the form: 
\be
y(\lambda,\sigma)=e^{-i\sigma/2}f(\lambda),
\ee
where a function $f(\lambda)$ is independent of $\sigma$ and satisfies
$f(0)=f(1)=0$. 

Integrals of the holomorphic two-form $\Omega$ on
$C_j$ become as follows:
\bea
\int_{C_j}\Omega
&=& \frac{i}{2}\int_{C_j}dv\wedge d\ln y(\lambda,\sigma) \nonumber\\
&=& \frac{1}{4}(e_{j+1}-e_j)\int_0^1d\lambda\int_0^{4\pi}d\sigma \nonumber\\
&=& \pi (e_{j+1}-e_j),
\eea
while integrals of the K\"ahler two-form are slightly entangled: 
\bea
\int_{C_j}\omega
&=&\frac{i}{4}\int_{C_j}
\frac{1}{V}\left(\frac{2dy}{y}-\delta dv\right)
\wedge\overline{\left(\frac{2dy}{y}-\delta dv\right)}\nonumber\\
&=&\frac{i}{4}\int_{C_j}
\frac{1}{V}\left\{Vdb+id\sigma+i{\rm Im}(\delta dv)\right\}
\wedge\left\{Vdb-id\sigma-i{\rm Im}(\delta dv)\right\}\nonumber\\
&=&\frac{1}{2}\int_{C_j}
\left\{db\wedge d\sigma+{\rm Im}(\delta dv)\wedge db\right\}\nonumber\\
&=&\frac{1}{2}\int_{C_j}db\wedge d\sigma\nonumber\\
&=&2\pi(b_{j+1}-b_j),
\eea
where we use the fact that ${\rm Im}(\delta dv)\wedge db$ vanish on $C_j$.

\newpage

%
%

\end{document}